# A Programmable and Reconfigurable Photonic Simulator for Classical XY Models

Jiayi Ouyang, Yuxuan Liao, Xue Feng,* Yongzhuo Li, Kaiyu Cui, Fang Liu, Hao Sun, Wei Zhang, and Yidong Huang
*Department of Electronic Engineering, Tsinghua University, Beijing 100084, China*

In this work, we proposed and experimentally demonstrated a photonic simulator for XY models, which is a typical kind of classical spin model. By encoding the XY spins on the phase term of the input light field, the corresponding XY Hamiltonian could be performed on the output light intensities. The simulator is mainly based on a programmable and reconfigurable optical vector-matrix multiplication system, which can map arbitrary XY models within the dimensionality limit. Here, we demonstrated the Berezinskii-Kosterlitz-Thouless transition in a two-dimensional XY model, in which the expectation values of some observables are calculated and consistent with the theory. Besides, we performed the ground state search of two 25-spin XY models with different spin connections and coupling strengths. Our proposal paves a new way to investigate the XY spin system.

Simulating the dynamics of classical spin models is an important issue in both physics and computer science [1]. The classical spin model consists of spins arranged in a lattice, in which the spin at site $i$ is denoted with a unit-length vector $\boldsymbol{S}_i$. Given the spin interaction matrix $\boldsymbol{J} = (J_{ij})$, the Hamiltonian of the spin system is $H = -\sum_{(i,j)} J_{ij}\boldsymbol{S}_i \cdot \boldsymbol{S}_j$. If the spins are planar rotors with $\boldsymbol{S}_i = (\cos\theta_i, \sin\theta_i)$, such model is known as an XY model [2]. The XY model can be employed in directional statistics [3], or associated with some important physical phenomena, including broken-symmetry transitions [2, 4], superfluid thin films [5], superconductors [6], etc.. Besides, searching the ground state of the XY model can be related to finding the optimal solution of the NP-hard continuous complex constant modulus quadratic optimization problem [2]. Recently, some physical platforms have been proposed to simulate the XY models, including those based on optical parametric oscillators (OPOs) [3], polaritions [2, 7, 8], laser networks [9], or on-chip phased arrays [10]. However, it is still challenging to implement arbitrary interactions among the XY spins. For instance, the spin modulation is time-multiplexed and the couplings are introduced with the optical delay line in the XY simulator on OPOs [3], in which implementing the all-to-all spin connections is difficult. In the polariton simulators [2, 7, 8], the spin interactions mainly depend on the separation distances among the polaritons. For the laser network approach, the aperture/lens is employed to achieve the weak/strong spin couplings [9]. Thus, for both the polariton simulators and the laser networks, it is hard to perform an XY model with arbitrary interaction matrix $\boldsymbol{J}$. Additionally, in the on-chip photonic phased array [10], the relation between the chip configuration and the XY model is still not clear and explicit. As shown by the simulators mentioned above, there is still no programmable XY simulator that can perform arbitrary XY models and various tasks. In this work, we experimentally implemented a programmable and reconfigurable photonic simulator that can demonstrate XY models with arbitrary coupling strengths and connections. Our proposed simulator is some kind of optoelectronic annealer based on the optical vector-matrix multiplication (OVMM) platform and the electronic feedback with heuristic algorithms. During each iteration, each spin $\boldsymbol{S}_i$ is encoded on the phase term of the light field and multiplied by a transformation matrix with the OVMM system. After the OVMM, the output intensity of the transformed light field would be detected by a photodetector. Then an electronic processor is utilized to compute the Hamiltonian from the detected optical intensity. Subsequently, it generates and updates the new spin configuration according to the employed heuristic algorithm for the next iteration. It should be mentioned that the employed OVMM is based on our previously proposed beam splitting and combining architecture [11–13], which can perform arbitrary complex vector-matrix multiplications. Meanwhile, different algorithms can be flexibly applied in the electronic domain to carry out different tasks. Thus, various tasks can be performed with our proposed programmable and reconfigurable photonic simulator. In the experiments, we observed the phenomena of the Berezinskii-Kosterlitz-Thouless (BKT) transition [14] in the two-dimensional (2D) XY model with 400 spins. Besides, the ground state search of two 25-spin XY models of different spin connections and coupling strengths was also presented. We believe that our method would provide a flexible and fast photonic system to simulate the classical XY models.

Fig. 1(a) shows a 2D XY model, where the spins are arranged in a square lattice. The spin at site $\boldsymbol{r}_i$, which is the translation vector of the spin $i$, possesses the azimuth angle $\theta_i \in [0, 2\pi)$ relative to the unit vector $\hat{\boldsymbol{x}}$, and can be denoted by the magnetization vector $\boldsymbol{S}_i = (\cos\theta_i, \sin\theta_i)$. The corresponding Hamiltonian can be obtained by summing all spin interactions [2, 15]:

$$H = -\sum_{(i,j)} J_{ij}\boldsymbol{S}_i\boldsymbol{S}_j = -\frac{1}{2}\boldsymbol{\Theta}^{\mathrm{H}}\boldsymbol{J}\boldsymbol{\Theta}, \qquad (1)$$

where $(i, j)$ denotes a pair of spins at site $\boldsymbol{r}_i$ and $\boldsymbol{r}_j$

with interaction strength $J_{ij}$, and the superscript H denotes the conjugate transpose. Eq. (1) indicates that the Hamiltonian can also be expressed with the complex quadratic form with the spin configuration vector $\boldsymbol{\Theta} = [\exp(\mathrm{i}\theta_1), \exp(\mathrm{i}\theta_2), \ldots]^{\mathrm{T}}$ (the superscript T denotes the transpose). To simulate such models, a photonic system is employed to encode the spin configuration on the light field and perform the spin interactions. As shown in Fig. 1(b), our proposed photonic XY simulator consists of three modules. The first is the beam-generation module that comprises a laser, a collimator, and a spatial light modulator (SLM). The SLM (SLM0 in Fig.1(b)) splits the beam from the collimator into $M$ beams, each of which will encode a spin. The second is the OVMM module with two SLMs (SLM1 and SLM2 in Fig.1(b)) and a pinhole. SLM1 encodes the azimuth angle $\theta_i$ of each spin on the phase term of each beam, which corresponds to the input vector $\boldsymbol{\Theta}$. Besides, SLM1 and SLM2 conduct a matrix transformation $\boldsymbol{A}$ by properly splitting and recombining the beams [13]. Then the pinhole filters out the beams with unwanted directions. The last is the detection module. A lens first aligns the beams to the optical axis, then a detector measures the output optical intensities of the OVMM, $\boldsymbol{I} = (\boldsymbol{A}\boldsymbol{\Theta})^{\mathrm{H}} \odot (\boldsymbol{A}\boldsymbol{\Theta})$ ($\odot$ denotes the element-wise production). The detailed setup is provided in Supplementary Material [16] (see also reference [12, 13, 15, 17–19] therein). We noticed that the Hamiltonian in Eq. (1) can be directly calculated from $\boldsymbol{I}$ by configuring the transformation matrix $\boldsymbol{A}$ satisfying $\boldsymbol{A}^{\mathrm{H}}\boldsymbol{A} = \boldsymbol{J} + \boldsymbol{J}_D$ [16]):

$$H(\boldsymbol{\Theta}) = -\frac{1}{2}(\boldsymbol{A}\boldsymbol{\Theta})^{\mathrm{H}}(\boldsymbol{A}\boldsymbol{\Theta}) + H_0 = -\frac{1}{2}\sum_i I_i + H_0, \quad (2)$$

where $\boldsymbol{J}_D$ is a diagonal matrix, and $H_0$ is a constant. As our primary concern is the relative Hamiltonian, the constant term $H_0$ is neglected in the following discussion.

In the operation process of the simulator, the OVMM module is first configured according to the transformation matrix $\boldsymbol{A}$, which is unchanged during the following process. A spin state $\boldsymbol{\Theta}$ is generated and encoded as the input vector. Then an electronic processor is employed to calculate the Hamiltonian from the output intensities with Eq. (2), and decide whether to accept the current spin state or not. The above procedure from generating a spin state to the acceptance of such a state is regarded as one sampling. The sampling should be conducted iteratively to collect enough statistics according to the specific objectives.

Due to the full programmability of the OVMM, our simulator can demonstrate XY models with arbitrary connections and coupling strength. Besides, the reconfigurability of our simulator allows us to configure different XY models and employ different algorithms to carry out different tasks. In the experiment, two tasks including observing the phase transition and searching the ground states are demonstrated with different XY models.

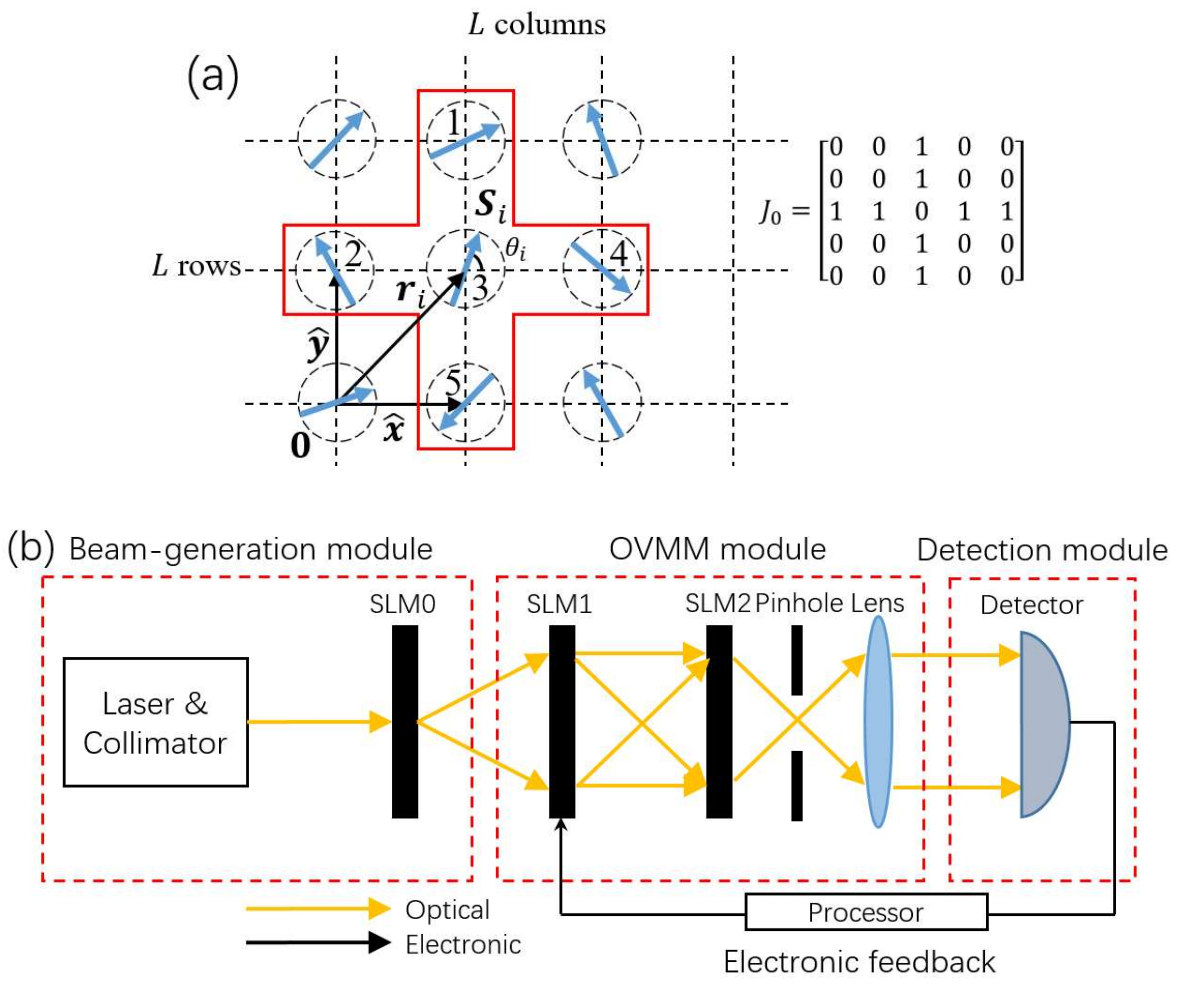

FIG. 1. Experimental setup of the photonic XY simulator. (a) A two-dimensional XY model. (b) Experimental setup of the photonic XY simulator. The detailed principle of the OVMM is provided in Supplementary Material [16]. Laser: ORION 1550nm Laser Module. SLM: Holoeye PLUTO-2.1-TELCO-013). Detector: Hamamatsu InGaAs Camera C12741-03.

First, our proposed photonic simulator is employed to demonstrate the BKT transition [14] by performing Boltzmann sampling of the 2D XY model under a series of temperature stages. The model has $N = 400$ spins in a square lattice ($L = 20$) with periodic boundary conditions. In this model, each spin only interacts with four nearest neighbors with the strength of $J_{ij} = 1$, as shown in Fig. 1(a). The Metropolis algorithm [20, 21] is employed for the Boltzmann sampling. In each sampling iteration, the angle of one random spin is changed to a random direction. Then the spin perturbation is accepted with the probability of $\min[1, \exp(-\Delta H/T)]$, where $\Delta H$ is the Hamiltonian variation resulting from the perturbation. The annealing temperature $T$ is the product of the Boltzmann constant and the real temperature for convenience. As the perturbation of a single spin involves $C = 5$ spins within the red box in Fig. 1(a), only the five spins, rather than the entire spins, are configured to the photonic simulator to calculate the Hamiltonian variation. Besides, the transformation matrix of the OVMM should be configured according to the five-spin interaction matrix $\boldsymbol{J}_0$ shown in Fig. 1(a). Since each site in the lattice is equivalent due to the periodic boundary condition, $\boldsymbol{J}_0$ and $\boldsymbol{A}$ are unchanged in the iterative sampling process. The maximum dimensionality of the employed OVMM is $M = 25$, hence our simulator can process $P = M/C = 5$ spin perturbations simultaneously. In each sampling iteration, $P$ nonadjacent spins are randomly selected. Each selected spin combined with its neighbors forms a spin group including $C = 5$ spins. Then $P$ spin groups are configured to the simulator, and the interaction energy $H_i^{(1)} (i = 1, 2, \ldots, P)$ of each group is calculated from the output intensities with Eq. (2). Then the $P$ selected spins are perturbed, and the new interaction energy $H_i^{(2)}$ $(i = 1, 2, \ldots, P)$ of each group is obtained. Each spin perturbation is accepted with the probability of $\min\{1, \exp[-(H_i^{(2)} - H_i^{(1)})/T]\}$. The accepted spin configurations are recorded for the following calculations.

The employed SLM has 8-bit grayscale so that the spins can only take angles of $2\pi n_i/q$ $(n_i = 0, 1, \ldots, q - 1, q = 256)$. Additionally, considering the sampling speed of the SLM and the camera, $q = 32$ is chosen for demonstrating the BKT transition [22]. The annealing temperature $T$ is slowly decreased from 2 to 0.2, including 19 stages with an interval of about 0.1. In each temperature stage, there are $2 \times 10^4$ sampling iterations, including $10^5$ spin perturbations. The last $5 \times 10^3$ samplings in each temperature stage are used to calculate three observables under the corresponding temperature, including the magnetic susceptibility $\chi = \left\langle [\sum_i \cos(\theta_i)]^2 \right\rangle / N$ [23], the spin correlation function $G(r) = \langle S(\boldsymbol{r}) S(\boldsymbol{0}) \rangle$ [14], and the helicity modulus [24]

$$\Upsilon = -\frac{\langle H \rangle}{2N^2} - \frac{1}{TN^2} \left\langle \left[ \sum_{i,j} \sin(\theta_i - \theta_j)\, \hat{\boldsymbol{e}}_{ij} \cdot \hat{\boldsymbol{x}} \right]^2 \right\rangle, \quad (3)$$

where $\langle \cdot \rangle$ represents the average value under a certain temperature, $\hat{\boldsymbol{e}}_{ij} = (\boldsymbol{r}_j - \boldsymbol{r}_i) / |\boldsymbol{r}_j - \boldsymbol{r}_i|$ is the unit translation vector from site $i$ to $j$, and $\hat{\boldsymbol{x}}$ is the unit vector of $x$ axis in the lattice plane as shown in Fig. 1(a). Fig. 2(a) shows the evolution of the magnetic susceptibility $\chi$. It can be observed that when temperature $T > 1.2$, $\chi$ only fluctuates within a small range around the mean value close to 0, which indicates that the spin system is highly disordered. When $T$ decreases under the value of 1.2, the mean value of $\chi$ rapidly grows and $\chi$ distributes within a broader range, which indicates the ferromagnetism of the system. The spin correlation function denotes the correlation between the spin directions at position $\boldsymbol{r}$ and position $\boldsymbol{0}$ as shown in Fig. 1a. The values of $G(r)$ $(r = |\boldsymbol{r}|)$ under different temperatures are illustrated in Fig. 2(b). It can be seen that $G(r)$ decays sharply when $T > 1.2$, indicating that the spin directions are highly uncorrelated. When $0.6 < T < 1.2$, the decay of $G(r)$ is slower. Suddenly, $G(r)$ decays very slowly when $T < 0.6$, which shows a quasi-long-range order. Such results are consistent with the theoretical predictions [14]. Thus, the system experienced a BKT transition: when $T$ is higher than the critical temperature $T_c$, $G(r)$ experiences a power-law decay, while $G(r)$ decays exponentially when $T < T_c$. To obtain the accurate critical temperature $T_c$, the evolution of the helicity modulus $\Upsilon$ is shown in Fig. 2(c), Fig. 2(c) shows that $\Upsilon$ fluctuates around 0 when $T$ decreases from 2 to 1.2, while $\Upsilon$ increases rapidly when $T$ further decreases to 0.2. The intersection of the measured curve and the line of $2T/\pi$ denotes the critical temperature $T_c \approx 0.91$. Besides, another evidence corresponding to the topological characteristic in the BKT transition is the vortices and anti-vortices [14]. To simulate such a phenomenon, another sampling is conducted at the temperature of around 0.1, which is much below $T_c$. After about 6000 iterations, the recorded spin orientations are shown in Fig. 2(d). It can be observed that there is a vortex (the red box) and an anti-vortex (the blue box), which is consistent with the theory [14].

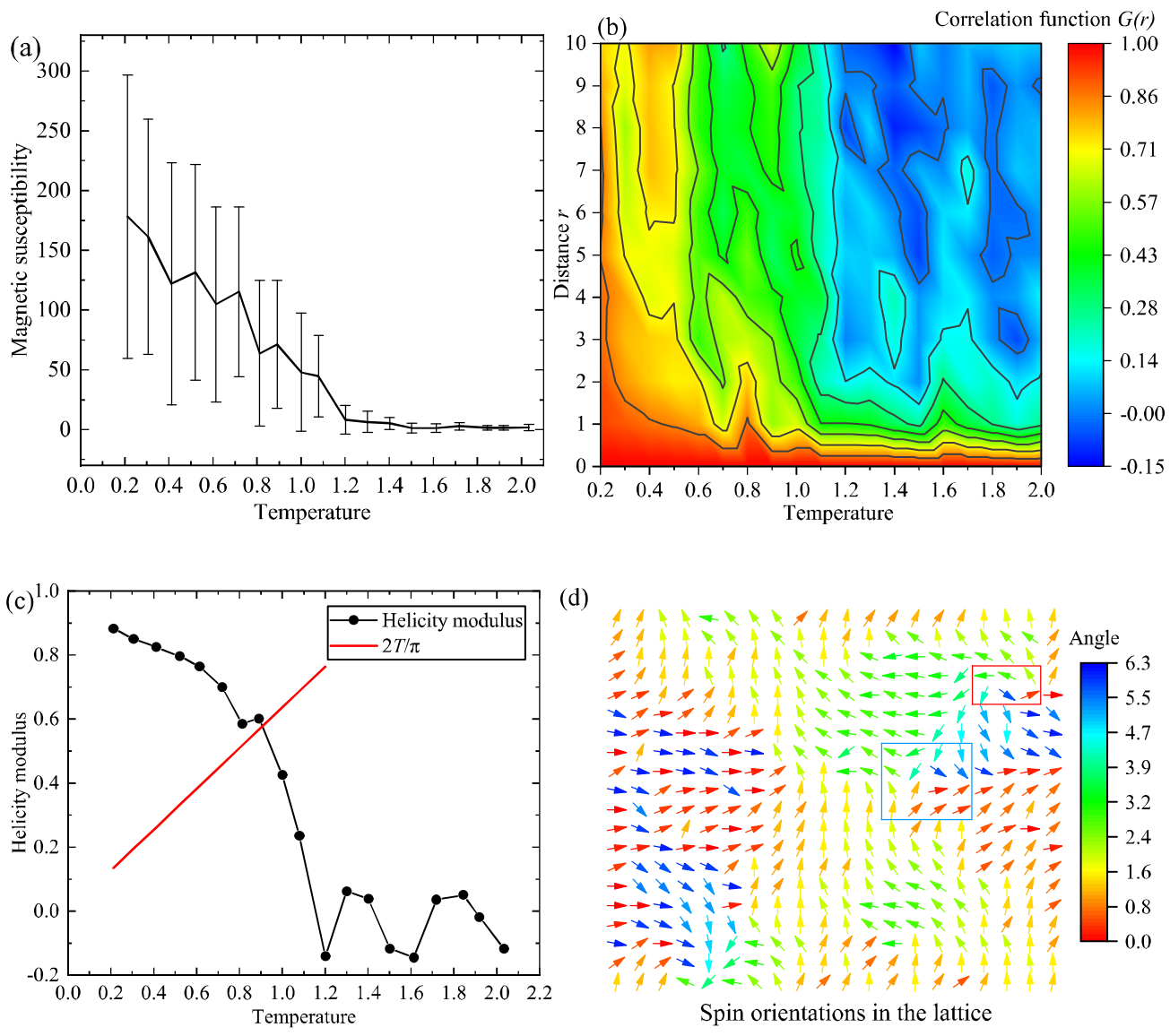

FIG. 2. The measured experimental observations of the 2D XY model. (a) The magnetic susceptibility versus the temperature. The error bar shows the standard deviation. (b) The correlation function versus distance and temperature. (c) The helicity modulus versus the temperature. The straight curve is $2T/\pi$ and the intersection of the two curves denotes the critical point. (d) Observation of a bound vortex-antivortex pair below the critical temperature. The red/blue box includes a vortex/antivortex.

Besides the sparse model, different XY models with arbitrary connections and coupling strength can also be performed with our proposed simulator. For the demonstration, the ground state search of two 25-spin XY models is performed. Here, the SLMs perform the maximum spin angle levels of $q = 256$, which is high enough to approximate an XY model. It should be mentioned that, when solving the XY models mapped from the optimization problems rather than the real spin systems, the connections and coupling strengths would be complicated, hence two models are randomly generated as follows. For model 1, the spin coupling strengths are uniformly distributed in $[-1, 1]$, and the graph density $2Q/[M(M-1)]$, ($Q$ is the number of the spin connections and $M$ is the number of spins) is 0.56 as shown in Fig. 3(a). Model 2 is fully connected with coupling strengths uniformly distributed in $\{-1, 1\}$, as shown in Fig. 3(b). The interaction matrices of models 1 and 2 are also provided in Supplementary Material [16]. When dealing with XY models with complex interactions, all 25 spins have to be implemented in the simulator simultaneously. According to Eq. (1), finding the ground state can be regarded as solving a continuous complex quadratic optimization problem [2, 15], which is generally NP-hard. Here, heuristic algorithms are employed to obtain the near-optimal solutions efficiently [25], and our adopted algorithm is the fast simulated annealing algorithm [18, 19]. In each iteration, a 25-dimensional Cauchy variable $\Delta\mathbf{\Theta}$ is added to the current spin configuration $\mathbf{\Theta}$. The new spin configuration $\mathbf{\Theta} + \Delta\mathbf{\Theta}$ is accepted with probability of $\min[1, \exp\{-[H(\mathbf{\Theta} + \Delta\mathbf{\Theta}) - H(\mathbf{\Theta})]/T\}]$. The temperature $T$ is gradually decreased and the near-optimal solutions are obtained finally. The experimental results are shown in Fig. 3. 50 runs of the ground state search are conducted for model 1/2 and the corresponding normalized Hamiltonian evolution curves are shown in Fig. 3(a)/3(b). It can be seen that all curves quickly converge to low Hamiltonians within about 1000/1500 iterations for model 1/2. To evaluate the searching performance, we consider one search to be successful when the final accepted Hamiltonian is lower than the lowest Hamiltonian (obtained in the simulation, $H_{\min} \approx -78.0/-193.8$ for model 1/2) multiplied by a tolerance coefficient $\eta$ since it is hard to find the exact global optimum. The successful probabilities under different tolerance coefficients (from 0.85 to 0.95) in each sampling iteration are calculated and shown in Fig. 3(c)/3(d) for model 1/2. The final successful probabilities are close to 1 when $\eta < 0.9$, but rapidly decrease when $\eta$ further increases to 0.95. For comparison, numerical simulations are conducted and the final successful probabilities for both models are 1 with $\eta = 0.95$. Such deterioration mainly results from the detection noise [13], even though the average fidelities of all effective samplings are as high as 0.9970±0.0033 and 0.9967±0.0031 for models 1 and 2, respectively [16]). Nevertheless, the results validate the capability of our photonic simulator to solve the XY models with complicated connections and interactions.

In this article, we proposed a programmable and reconfigurable photonic XY simulator, which can demonstrate XY models with arbitrary interactions and different tasks. After encoding the spin configuration on the phase term of the light field, our simulator can provide the corresponding Hamiltonian within the propagation time of light beam. To validate our proposal, we demonstrate the BKT transition of the 2D XY model and perform the ground state search of two models with different connections and coupling strengths.

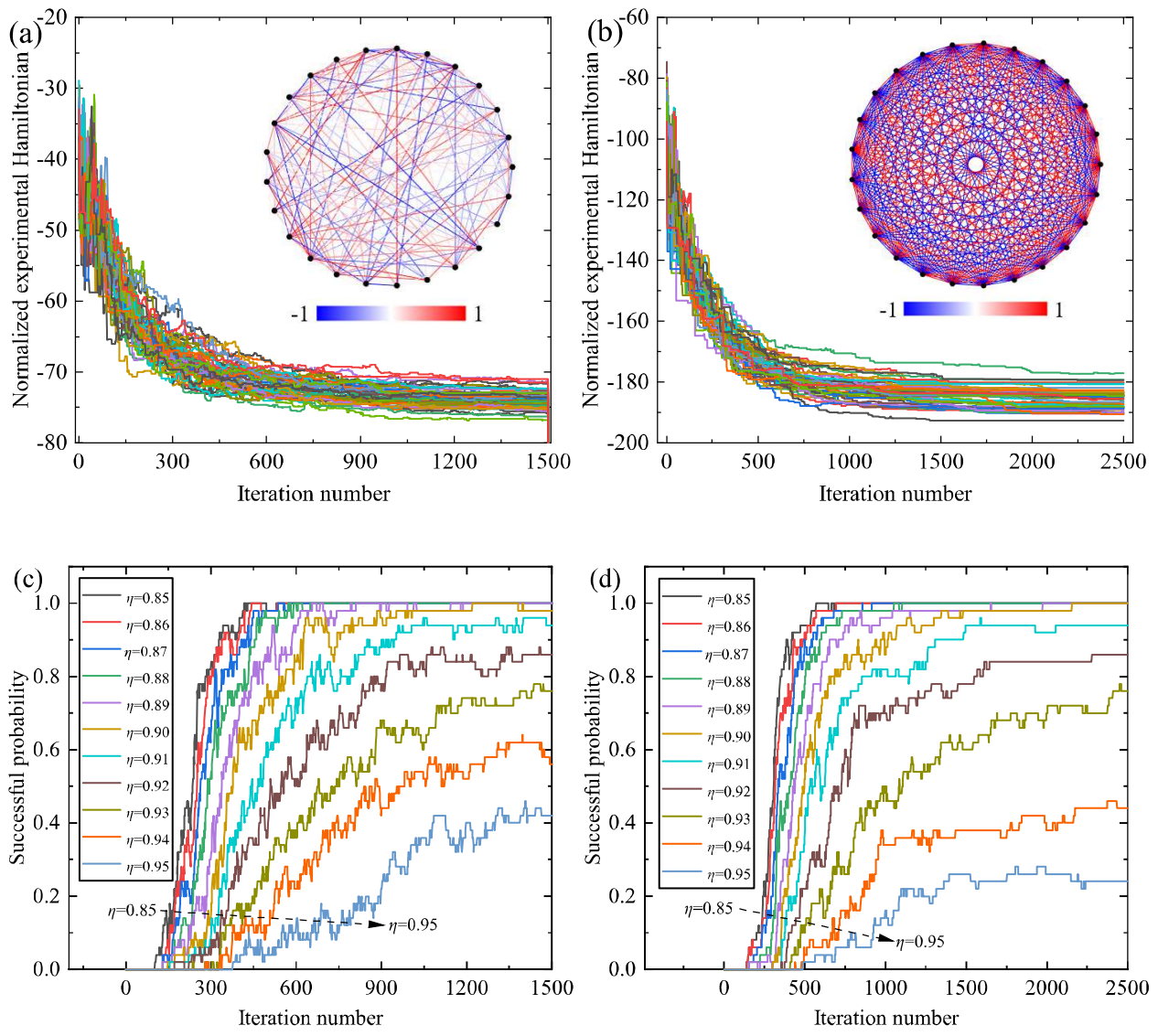

FIG. 3. Results of the ground state search of the randomly generated XY models. (a) (b) 50 normalized experimental Hamiltonian evolution curves for models 1 and 2 respectively. The insets show the corresponding models, where the black dots denote the XY spins and the interaction strengths are referred from the colorbar. (c) (d) The successful probabilities versus the iteration number under different tolerance coefficients $\eta$ in the experiment of model 1 and 2 respectively. The black dashed arrow successively intersects the curves with growing $\eta$ from 0.85 to 0.95 in each figure.

Two major issues can be further improved for our proposal. The first is to increase the dimensionality of the OVMM, which is helpful to process more spin perturbations simultaneously in sparse models, or deal with complex models with higher dimensionalities. In our employed OVMM, increasing the dimensionality could be achieved by reducing the size of the Gaussian beam (constrained by paraxial approximation) and employing SLMs with larger areas. Besides, other optical computation schemes could also be utilized to implement our photonic simulator, such as the on-chip Mach–Zehnder interferometer arrays [26, 27].

The second is the time consumption of the optoelectronic conversion in the SLMs and camera, which limits the computation speed of our simulator. In the future, the utilization of the high-speed phase modulators [28] and detectors [29] would improve the state-sampling speed in the optical domain. In the electronic domain, the speed can be increased by employing an FPGA, or analog computing devices [30].

This work was supported by the National Key

Research and Development Program of China (2018YFB2200402, 2017YFA0303700), and the National Natural Science Foundation of China (Grant No. 61875101). This work was also supported by Beijing academy of quantum information sciences, Beijing National Research Center for Information Science and Technology (BNRist), Frontier Science Center for Quantum Information, and Tsinghua Initiative Scientific Research Program.

* x-feng@tsinghua.edu.cn

Supplementary Information for

# A Programmable and Reconfigurable Photonic Simulator for Classical XY models

Jiayi Ouyang, Yuxuan Liao, Xue Feng*, Yongzhuo Li, Kaiyu Cui, Fang Liu, Hao Sun, Wei Zhang, and Yidong Huang

Department of Electronic Engineering, Tsinghua University, Beijing 100084, China
*Correspondence: x-feng@tsinghua.edu.cn

**This file includes:**
Supplementary Notes 1 to 6
Figs. S1 to S5
References

**Supplementary Note 1: Relation between the classical XY Hamiltonian and the experimental parameters**

In this section, the method of calculating XY Hamiltonians with the experimental parameters of the photonic simulator is discussed in detail. In this discussion, only XY models in absence of external fields are considered, and the Hamiltonian of the $N$-spin XY model is:

$$H = -\frac{1}{2}\sum_{i,j} J_{ij}\boldsymbol{S}_i \cdot \boldsymbol{S}_j = -\frac{1}{2}\sum_{i,j} J_{ij}\cos(\theta_i - \theta_j), \tag{S1}$$

where $\boldsymbol{J} = (J_{ij})$ is the interaction matrix and $\boldsymbol{S}_i = (\cos\theta_i, \sin\theta_i)$ denotes a two-dimensional, unit-length vector. For simplicity, we assume that $\boldsymbol{J}$ is a real symmetric matrix (the asymmetric $\boldsymbol{J}$ can be converted to a symmetric matrix $(\boldsymbol{J} - \boldsymbol{J}^{\mathrm{T}})/2$ without affecting the Hamiltonian). The XY Hamiltonian can also be expressed with the quadratic form [1]:

$$H = -\frac{1}{2}\sum_{i,j} J_{ij}\cos(\theta_i - \theta_j) = -\frac{1}{2}\sum_{\langle i,j\rangle} J_{ij}\left(\exp(\mathrm{i}(\theta_i - \theta_j)) + \exp(-\mathrm{i}(\theta_i - \theta_j))\right) = -\frac{1}{2}\boldsymbol{\Theta}^{\mathrm{H}}\boldsymbol{J}\boldsymbol{\Theta}, \tag{S2}$$

where $\boldsymbol{\Theta} = [\exp(\mathrm{i}\theta_1), \exp(\mathrm{i}\theta_2), \ldots, \exp(\mathrm{i}\theta_N)]^{\mathrm{T}}$ is the complex spin configuration vector, and the superscripts H and T denote the conjugate transpose and the transpose respectively. By using the eigen-decomposition, the real symmetric matrix $\boldsymbol{J}$ can be expressed as:

$$\boldsymbol{J} = \boldsymbol{U}^{\mathrm{H}}\boldsymbol{\Lambda}\boldsymbol{U}, \tag{S3}$$

where $\boldsymbol{\Lambda} = \mathrm{diag}(\lambda_1, \lambda_2, \ldots, \lambda_N)$ is the diagonal eigenvalue matrix with $N$ eigenvalues $\lambda_1, \lambda_2, \ldots, \lambda_N$, and $\boldsymbol{U}$ is the unitary eigenvector matrix composed by eigenvectors $\boldsymbol{u}_1, \boldsymbol{u}_2, \ldots, \boldsymbol{u}_N$. Before the next operation, we have to check whether $\boldsymbol{J}$ is positive semi-definite, which means all of its eigenvalues are non-negative:

**Step (1).** If the lowest eigenvalue of $\boldsymbol{J}$ is non-negative, skip step (2).

**Step (2).** If the lowest eigenvalue of $\boldsymbol{J}$ is negative, modify the diagonal elements of $\boldsymbol{J}$ to ensure that all of its eigenvalues are non-negative. Such modification of $\boldsymbol{J}$ will only introduce a global offset of the XY Hamiltonian, and will not affect the relative Hamiltonian. Here we list two methods of modifying $\boldsymbol{J}$:

(i) Replace each original diagonal element $J_{ii}$ by $\sum_{j,i\neq j}|J_{ij}|$ to ensure the new interaction matrix $\boldsymbol{J}'$ is diagonally dominant [2].

(ii) Denote the lowest eigenvalue by $\lambda_{\min}$. Define the new interaction matrix $\boldsymbol{J}' = \boldsymbol{J} + |\lambda_{\min}|\boldsymbol{I}$. Since $\boldsymbol{J}'\boldsymbol{u}_i = (\boldsymbol{J} + |\lambda_{\min}|\boldsymbol{I})\boldsymbol{u}_i = (\lambda_i + |\lambda_{\min}|)\boldsymbol{u}_i$, $\boldsymbol{J}'$ has the non-negative eigenvalues $\lambda_1 + |\lambda_{\min}|, \lambda_2 + |\lambda_{\min}|, \ldots, \lambda_N + |\lambda_{\min}|$ and eigenvectors $\boldsymbol{u}_1, \boldsymbol{u}_2, \ldots, \boldsymbol{u}_N$.

In the following discussion, we use $\boldsymbol{J}$, $\boldsymbol{\Lambda}$, and $H$ to denote the new interaction matrix, eigenvalue matrix, and Hamiltonian after modifying the original interaction matrix respectively for simplicity. According to Eq. (S2),

$$H(\boldsymbol{\Theta}) = -\frac{1}{2}\boldsymbol{\Theta}^{\mathrm{H}}\boldsymbol{U}^{\mathrm{H}}\sqrt{\boldsymbol{\Lambda}}\sqrt{\boldsymbol{\Lambda}}\boldsymbol{U}\boldsymbol{\Theta} = -\frac{1}{2}\left(\sqrt{\boldsymbol{\Lambda}}^{\mathrm{H}}\boldsymbol{U}\boldsymbol{\Theta}\right)^{\mathrm{H}}\left(\sqrt{\boldsymbol{\Lambda}}\boldsymbol{U}\boldsymbol{\Theta}\right). \tag{S4}$$

Since $\lambda_1, \lambda_2, \ldots, \lambda_N \geq 0$, $\sqrt{\boldsymbol{\Lambda}}^{\mathrm{H}} = \sqrt{\boldsymbol{\Lambda}}$. By introducing $\boldsymbol{A} = \sqrt{\boldsymbol{\Lambda}}\boldsymbol{U}$ the XY Hamiltonian can be expressed as

$$H(\boldsymbol{\Theta}) = -\frac{1}{2}(\boldsymbol{A}\boldsymbol{\Theta})^{\mathrm{H}}(\boldsymbol{A}\boldsymbol{\Theta}). \tag{S5}$$

Therefore, we set the transform matrix to $\boldsymbol{A}$ and the input vector to $\boldsymbol{\Theta}$ in the optical vector-matrix multiplication (OVMM). The output field is $\boldsymbol{E} = \boldsymbol{A}\boldsymbol{\Theta} = [E_1, E_2, \ldots, E_N]^{\mathrm{T}}$, and the measured intensity vector is

$$\boldsymbol{I} = \boldsymbol{E}^{*} \odot \boldsymbol{E} = [|E_1|^2, |E_2|^2, \ldots, |E_N|^2]^{\mathrm{T}} = [I_1 \quad \cdots \quad I_k \quad I_{k+1} \quad \cdots \quad I_N]^{\mathrm{T}}, \tag{S6}$$

where the superscript * denotes the complex conjugation, and $\odot$ denotes element-wise multiplication. Thus, the XY Hamiltonian could be expressed with

$$H(\boldsymbol{\Theta}) = -\frac{1}{2}\boldsymbol{E}^{\mathrm{H}}\boldsymbol{E} = -\frac{1}{2}\left(\sum_i I_i\right), \tag{S7}$$

In conclusion, in the pretreatment stage, the required transformation matrix $\boldsymbol{A}$ is obtained by the eigen-decomposition of the modified interaction matrix $\boldsymbol{J}$ and the Hamiltonian can be calculated by summing the measured intensities of the output field.

**Supplementary Note 2: Experimental setup of the photonic simulator**

Fig. S1 shows the experimental setup of the photonic simulator for the classical XY model, which mainly comprises a laser module (ORION 1550nm Laser Module), three spatial light modulators (SLMs, HOLOEYE PLUTO-2.1-TELCO-013), a camera (Hamamatsu InGaAs Camera C12741-03), and a processor. Fig. S2 shows the principle of the optical vector-matrix multiplication (OVMM) system. The laser beam from the collimator is incident on the OVMM based on SLMs. The pattern on SLM0 shown in Fig. S3(a) splits the incident beam to $M$ beams ($M$ is the dimensionality of the transformation matrix). Here the beam-splitting ratio represents the ratio of the complex amplitudes of the $M$ beams. For SLM0, the complex beam-splitting ratio is the same as the target input vector denoted by $\boldsymbol{B} = [b_1, b_2, \dots, b_M]^{\mathrm{T}}$, as shown in Fig. S2. The $M$ beams then incident on $M$ different regions of SLM1 shown in Fig. S3(b) respectively. Likewise, each region on SLM1 splits the incident beam to $M$ different regions on SLM2 shown in Fig. S3(c) respectively, and the complex splitting ratio of each region corresponds to each column of the transformation matrix $\boldsymbol{A} = (a_{ij})$ respectively. For instance, the region $i$ on SLM1 splits the incident beam $i$ to $M$ beams with complex amplitudes of $a_{i1}b_i, a_{i2}b_i, \dots, a_{iM}b_i$ respectively, as shown in Fig. S2. Then the region $i$ on SLM2 equally recombines the $M$ incident beams from the $M$ regions on SLM1 to a same direction on the output plane, resulting in an output amplitude of $c_i = \sum_{j=1}^{M} a_{ij}b_j$, as shown in Fig. S2. Each SLM pattern is optimized with a gradient-descent algorithm from the superposition of a series of blazed gratings, and the detailed pattern generation method is provided in the supplementary information of our previous work [3]. As mentioned above, the input vector is encoded on SLM1 by appending phase delays equal to the spin angles to the corresponding regions respectively. Besides, since our employed SLMs are reflective, a blazed grating is appended to each SLM to concentrate the intensity in the first diffraction order. Finally, a camera (Hamamatsu InGaAs Camera C12741-03) detects the output optical intensities, which will be used to calculate the Hamiltonian corresponding to the spin configuration with the processor. Besides, the processor generates the next spin configuration according to the employed algorithm, and updates it to SLM1 iteratively until the algorithm terminates. The distances between the devices are: SLM0 to SLM1: 0.820m; SLM1 to SLM2: 0.754m; SLM2 to pinhole: 0.373m.

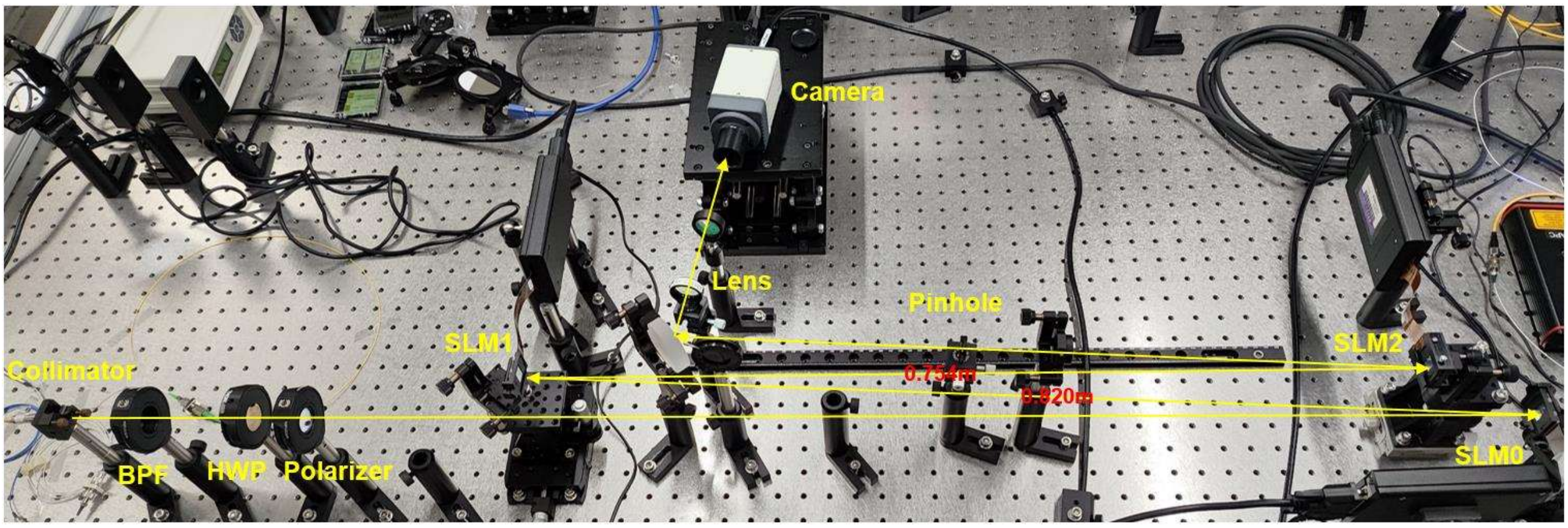

Fig. S1. The experimental setup of the photonic simulator. BPF: bandpass filter. HWP: half-wave plate.

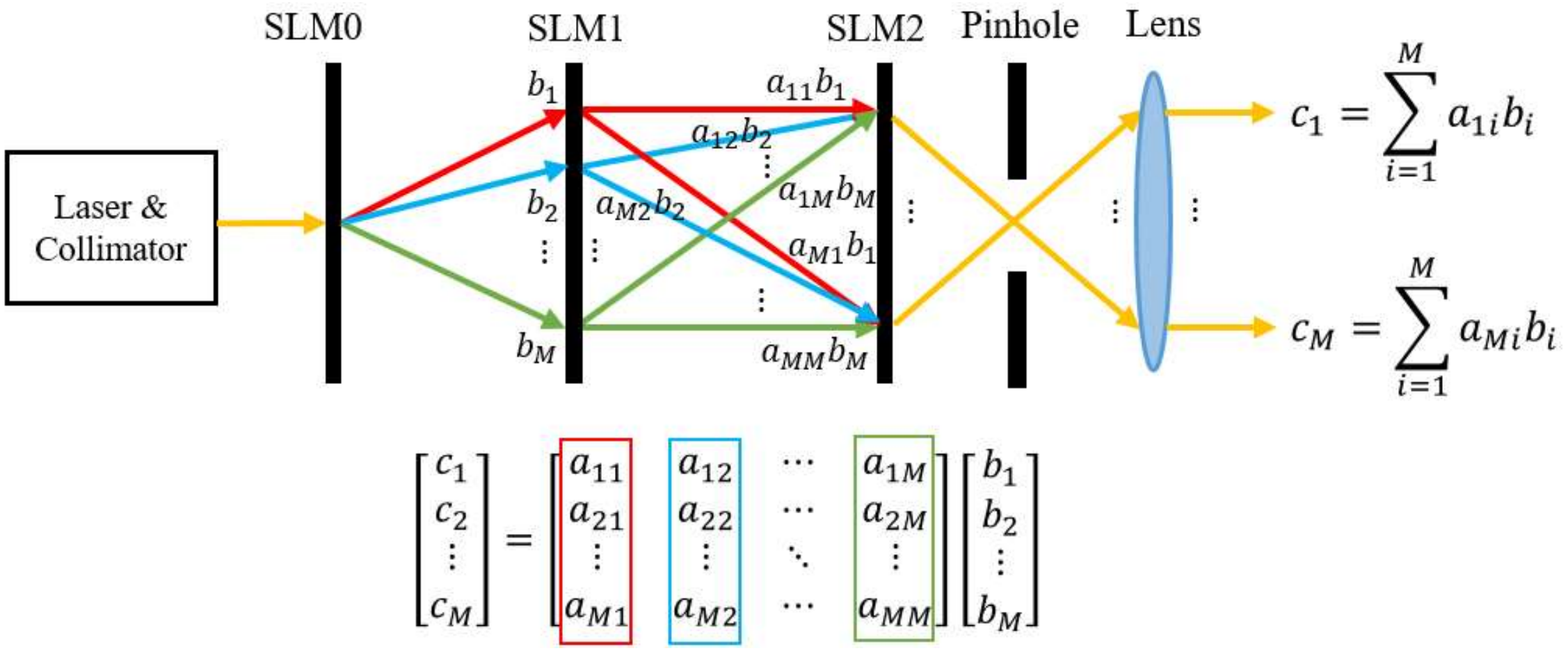

Fig. S2. The principle of the OVMM. The SLM1 splits each incident beam according to the corresponding column of the target transformation matrix denoted by the box with the same color, respectively. SLM2 then recombines the beams equally.

(a)

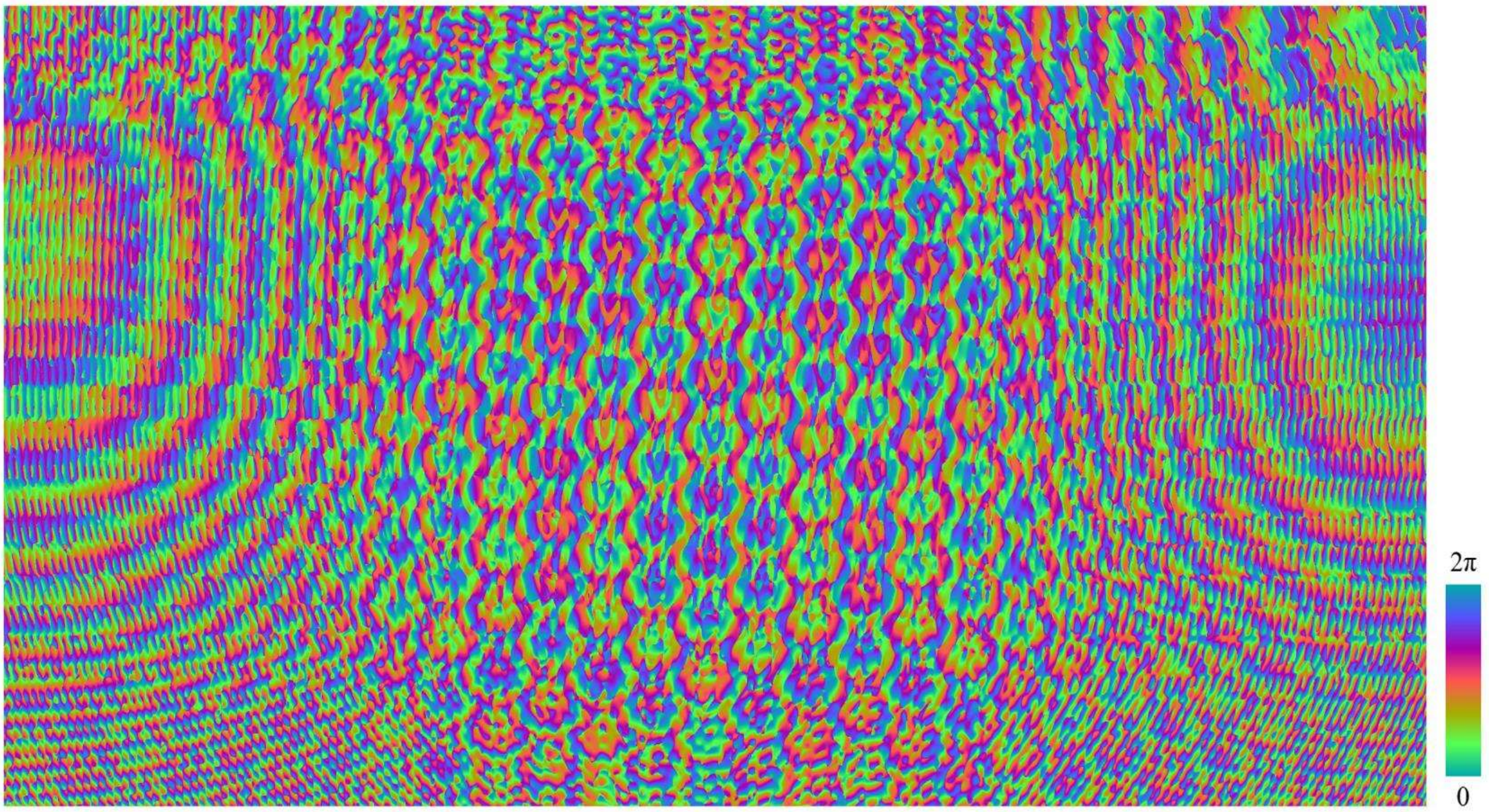

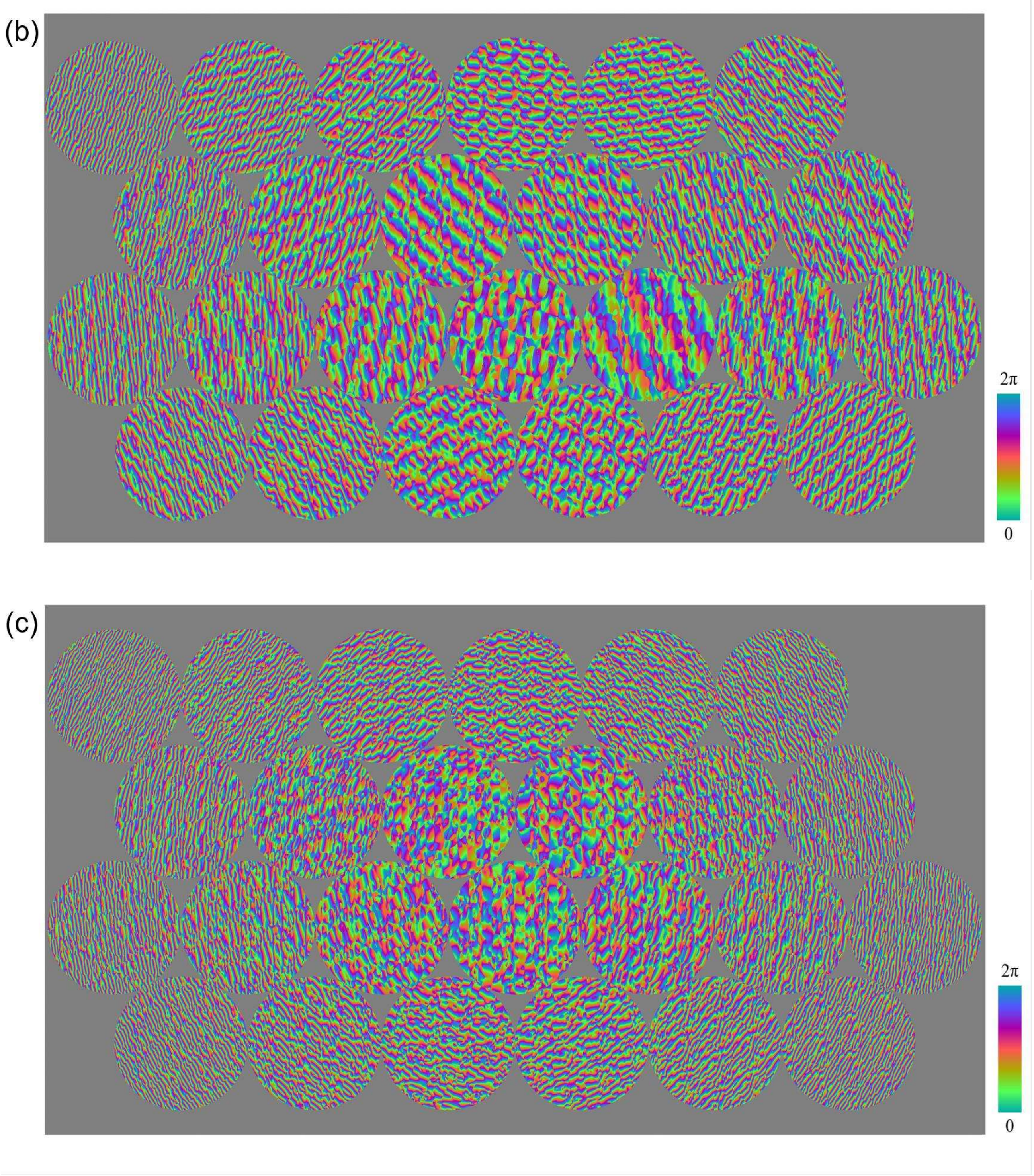

Fig. S3. The experimental SLM patterns for the model 2 when not appending the blazed grating. (a) The pattern on SLM0. (b) The pattern on SLM1. (c) The pattern on SLM2.

**Supplementary Note 3: Calibration of the SLM phase patterns for OVMM**

When employing the phase patterns generated from the ideal conditions, the actual transformation matrix would deviate from the target transformation matrix due to the misalignment and paraxial approximations. Therefore, the calibrations of both amplitude and phase terms are essential to reduce the systematic error. In this section, we use $A^{\mathrm{in}}$ and $A^{\text{out}}$ to denote the matrix input to the pattern-generation algorithm and the actual transformation matrix of the OVMM respectively, and

$$A_{ij}^{\text{in}} = A_{ij}^{\text{out}} \exp(\mathrm{i}\epsilon_{ij}), \tag{S8}$$

where the phase error $\epsilon_{ij}$ is independent of $A^{\text{in}}$. To measure the phase error, one of the $M$ beams incident on SLM1 is used as a reference beam (denoted as beam 1) to interference with one of the rest of the beams (denoted as beam $j$, $j = 2,3,\dots,M$) separately. The element $i$ $(i = 1,2,\dots,M)$ of the output intensity vector is

$$I_{ij}^{\cos} = \left|A_{i1}^{\text{out}} \exp(\mathrm{i}\epsilon_{i1}) + A_{ij}^{\text{out}} \exp(\mathrm{i}\epsilon_{ij})\right|^2. \tag{S9}$$

Then we use beam 1 to interfere beam $j$ with phase delay $\pi/2$:

$$I_{ij}^{\sin} = \left|A_{i1}^{\text{out}} \exp(\mathrm{i}\epsilon_{i1}) + \mathrm{i}A_{ij}^{\text{out}} \exp(\mathrm{i}\epsilon_{ij})\right|^2. \tag{S10}$$

It is reasonable to set the phase of the reference beam $\epsilon_{i1} = 0$. Now we can obtain the phase compensation

$$\epsilon_{ij} = \begin{cases} -\arccos\left[\dfrac{I_{ij}^{\cos} - (A_{i1}^{\text{out}})^2 - (A_{ij}^{\text{out}})^2}{2A_{i1}^{\text{out}}A_{ij}^{\text{out}}}\right], I_{ij}^{\sin} - (A_{i1}^{\text{out}})^2 - (A_{ij}^{\text{out}})^2 \geq 0 \\ \arccos\left[\dfrac{I_{ij}^{\cos} - (A_{i1}^{\text{out}})^2 - (A_{ij}^{\text{out}})^2}{2A_{i1}^{\text{out}}A_{ij}^{\text{out}}}\right], I_{ij}^{\sin} - (A_{i1}^{\text{out}})^2 - (A_{ij}^{\text{out}})^2 < 0 \end{cases}, \tag{S11}$$

where $A_{ij}^{\text{out}}$ can be obtained from the output intensities when each beam exists alone. By modifying the input matrix by $A_{ij}^{\text{in}} \exp(-\mathrm{i}\epsilon_{ij})$, the phase error can be well compensated.

The amplitude calibration can be conducted by $A_{ij}^{\text{in}} \to A_{ij}^{\text{in}}|A_{ij}^{\text{target}}|/A_{ij}^{\text{out}}$ repeatedly. The detailed procedure of the calibration method is provided in the supplementary information of the previous work [3,4].

**Supplementary Note 4: Fidelities of the transformation matrices of the OVMM**

In this section, the fidelities of the transformation matrices are provided. Fig. S4(a)-(c) show the interaction matrices $\boldsymbol{J}$ of the two-dimensional (2D) XY model, and model 1-2 respectively. It should be mentioned that the matrices shown in Fig. S4(a)-(c) are the modified positive semi-definite matrices according to Supplementary Note 1. Fig. S4(d)-(f) show the transformation matrices $\boldsymbol{A}$ of the OVMM corresponding to Fig. S4(a)-(c), respectively. To measure the actual experimental transformation matrix, the input vector is set to each column of the identity matrix successively, and then the output vectors form the output intensity matrix $\boldsymbol{I}_E$. Fig. S4(g)-(i) show the theoretical output intensity matrices $\boldsymbol{I}_T = \boldsymbol{A} \odot \boldsymbol{A}^*$, while Fig. S4(j)-(l) show the $\boldsymbol{I}_E$ corresponding to Fig. S4(d)-(f), respectively. It can be seen that the $\boldsymbol{I}_T$ and the $\boldsymbol{I}_E$ are very close to each other in each model. According to the definition of the fidelity of the transformation matrix [3]

$$f_{\text{mat}} = \frac{\sum_{ij}(\boldsymbol{I}_E)_{ij}(\boldsymbol{I}_T)_{ij}}{\sqrt{\sum_{ij}(\boldsymbol{I}_E)_{ij}^2}\sqrt{\sum_{ij}(\boldsymbol{I}_T)_{ij}^2}}, f_{\text{mat}} \in [0,1], \tag{S12}$$

we obtain the high matrix fidelities of 0.99997, 0.99887, and 0.99880 in the experiments of the 2D XY model, and model 1-2 respectively.

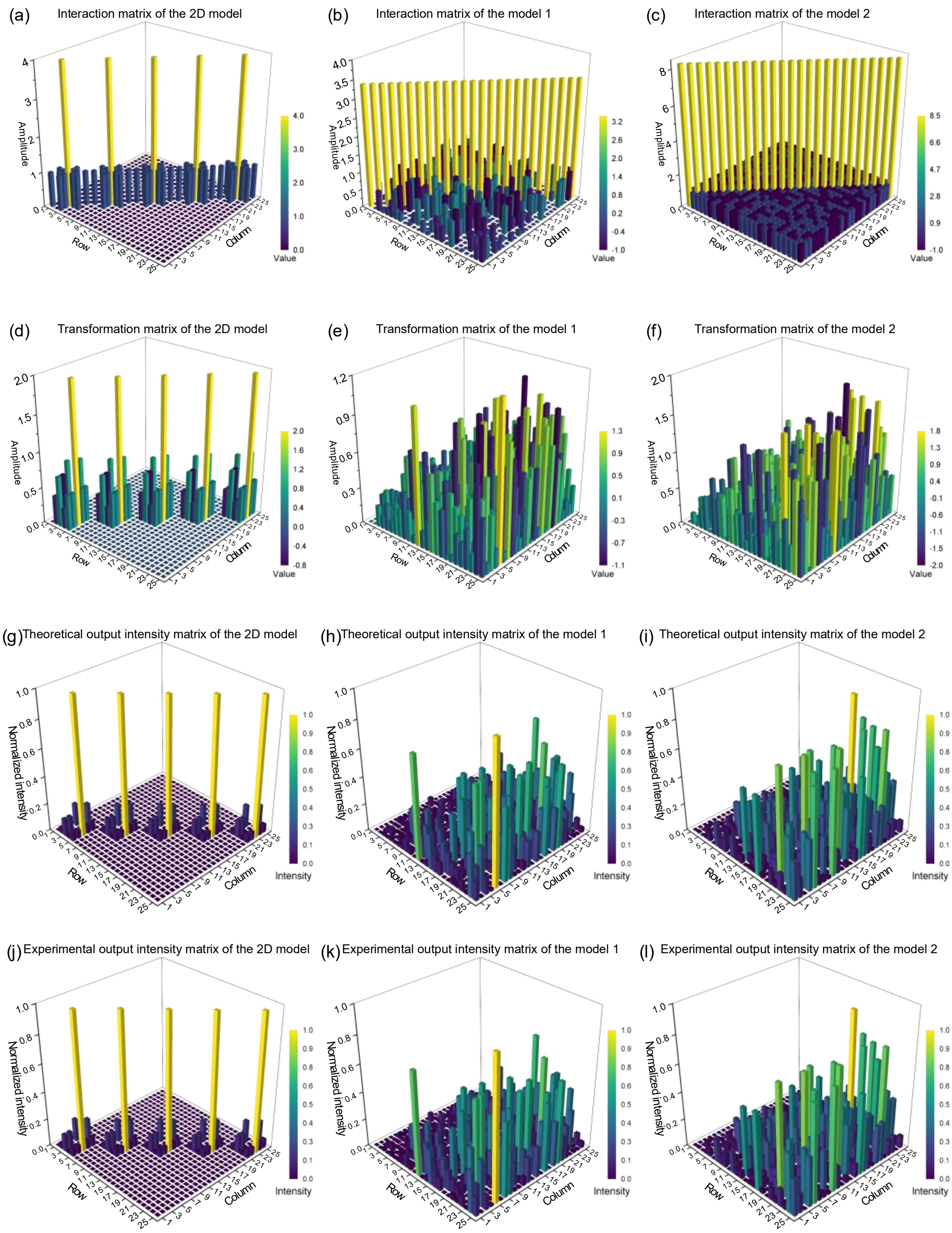

Fig. S4. The interaction matrices, transformation matrices, theoretical output intensity matrices and experimental output intensity matrices of the demonstrated XY models. (a)-(c) The interaction matrices of the 2D XY model, and model 1-2 respectively. (d)-(f) The transformation matrices of the OVMM of the 2D XY model, and model 1-2 respectively. (g)-(i) The theoretical output intensity matrices of the 2D XY model, and model 1-2 respectively. (j)-(l) The experimental output intensity matrices of the 2D XY model, and model 1-2 respectively.

**Supplementary Note 5: The accuracy of the output intensity vectors of the OVMM**

Besides the transformation matrix, the accuracy of the output intensity vectors is also investigated with two parameters. The first parameter is the fidelity of the intensity vector defined by [3]

$$f_{\text{vec}} = \frac{\sum_i (\boldsymbol{I}_{\text{exp}})_i (\boldsymbol{I}_{\text{theo}})_i}{\sqrt{\sum_i (\boldsymbol{I}_{\text{exp}})_i^2 \sum_i (\boldsymbol{I}_{\text{theo}})_i^2}}, f_{\text{vec}} \in [0,1], \tag{S13}$$

where $\boldsymbol{I}_{\text{exp}}$ denotes the experimental output intensity vector, while $\boldsymbol{I}_{\text{theo}}$ denotes the theoretical one. The more the two vectors are parallel, the more $f_{\text{vec}}$ is close to 1. The second parameter is the intensity stability parameter $K$, which is defined by $\sum_i (\boldsymbol{I}_{\text{exp}})_i / \sum_i (\boldsymbol{I}_{\text{theo}})_i$ and expected to be a constant. Due to the non-idealities and the detection noise in the experiment, $f_{\text{vec}} < 1$ and $K$ distributes within a small range. Fig. S5(a)-(c) show the average fidelities of all samplings in each temperature stage of the 2D model and in each run of model 1-2, respectively. The overall average fidelities in Fig. S5(a)-(c) are 0.9964±0.0039, 0.9970±0.0033, and 0.9967±0.0031, respectively. Fig. S5(d)-(f) show the average values of $K$ of all samplings in each temperature stage of the 2D model and in each run of model 1-2, respectively. The overall average $K$ in Fig. S5(d)-(f) are $(2.736\pm0.104)\times10^3$, $(1.472\pm0.055)\times10^2$, and 85.38±2.39 respectively, and the relative deviations are 3.8%, 3.7% and 2.8% respectively. The high vector fidelities and high stability of $K$ indicate that the OVMM in our photonic simulator is quite accurate.

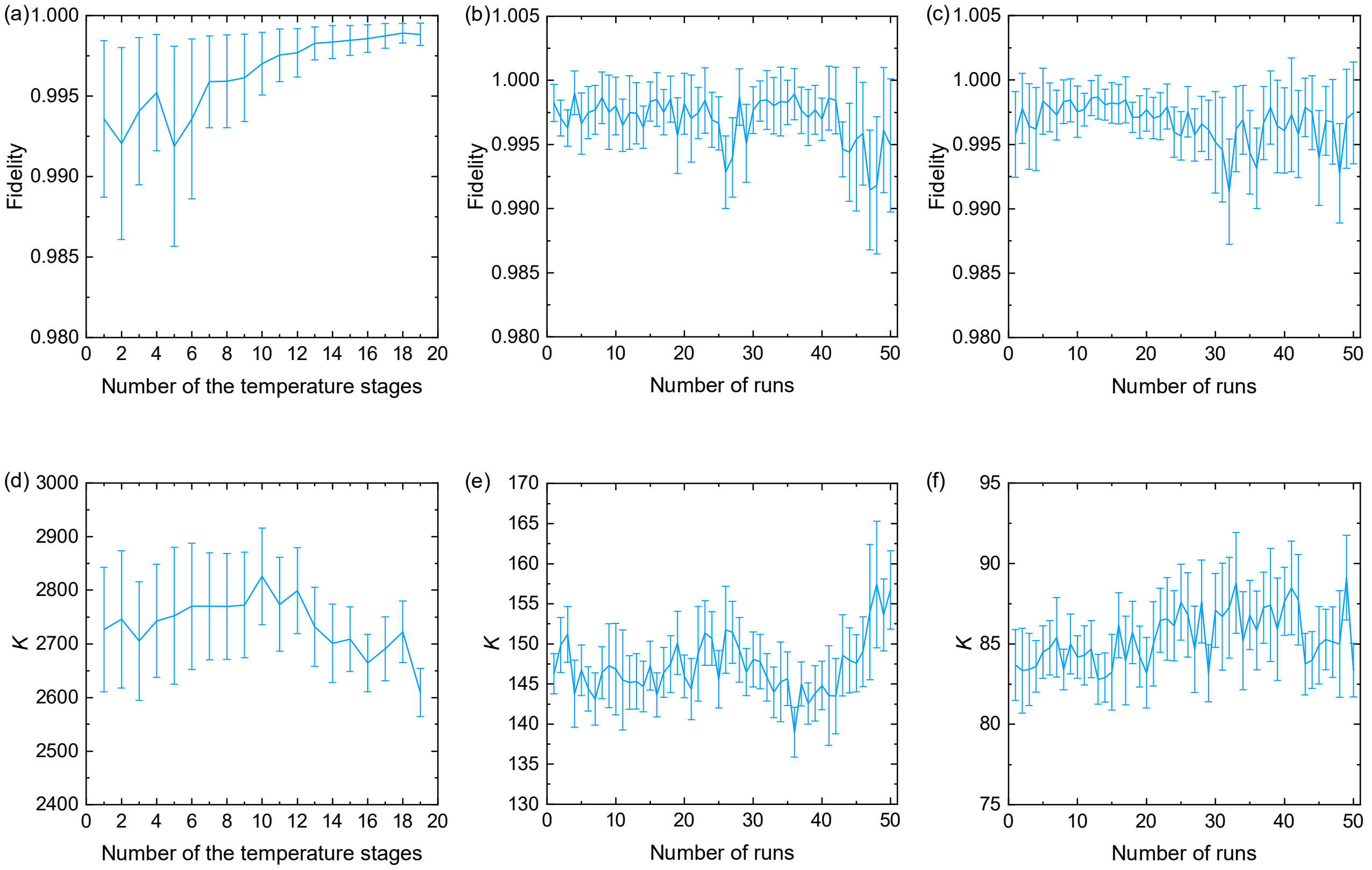

Fig. S5. The results of the average fidelities of the output intensity vectors and the average values of the intensity stability parameter $K$. (a)-(c) The average fidelities of the output intensity vectors of the 2D model and model 1-2 respectively. (d)-(f) The average values of the intensity stability parameter $K$ of the 2D model and model 1-2 respectively.

**Supplementary Note 6: The normalization of the experimental Hamiltonians and the successful probability of the ground state search**

In the experiment, the raw Hamiltonians $H_{\mathrm{exp}}$ of the samplings are calculated from the measured optical intensities. Fig. S6(a)/(b) show the 50 raw accepted Hamiltonian curves of the ground state search of model 1/2, respectively. In Supplementary Note 5, the average values intensity stability parameter $K$ of each run is calculated. Thus, the normalized experimental Hamiltonian $H_{\mathrm{n}}$ can be obtained by $H_{\mathrm{n}} = H_{\mathrm{exp}}/K$. It should be mentioned that, the theoretical annealing temperature of the 2D model is obtained with $T = T_{\mathrm{exp}}/K$, where $T_{\mathrm{exp}}$ is the annealing temperature in the experiment. The 50 curves in Fig. S6(a)/(b) are normalized according to their own average $K$ respectively, and the results are shown in Fig. S6(c)/(d).

Actually, we are concerned about whether the theoretical Hamiltonians of the accepted spin states are close enough to the ground state Hamiltonians. Therefore, the theoretical Hamiltonians of the accepted states are calculated, and the 50 theoretical Hamiltonian curves of model 1/2 are shown in Fig. S6(e)/(f). As mentioned in the main text, we consider a state sampling to be successful when its accepted theoretical Hamiltonian is lower than the ground state Hamiltonian (obtained with fast simulated annealing with the same parameters [5,6]) multiplied by a tolerance coefficient $\eta$. Thus, we calculated the successful probabilities of each iteration, which denotes the proportion of the successful samplings in 50 runs, under different $\eta$ from 0.85 to 0.95 with the interval of 0.01. The successful probability curves of model 1/2 are shown in Fig. S6(g)/(h). From these results, it can seen that the final successful probabilities are almost 1 when $\eta \leq 0.9$, while they quickly decay when $\eta > 0.9$. Such decay mainly results from the detection noise of the camera [3], which would make our photonic simulator unable to distinguish two spin configurations when their Hamiltonian difference is lower than the noise level.

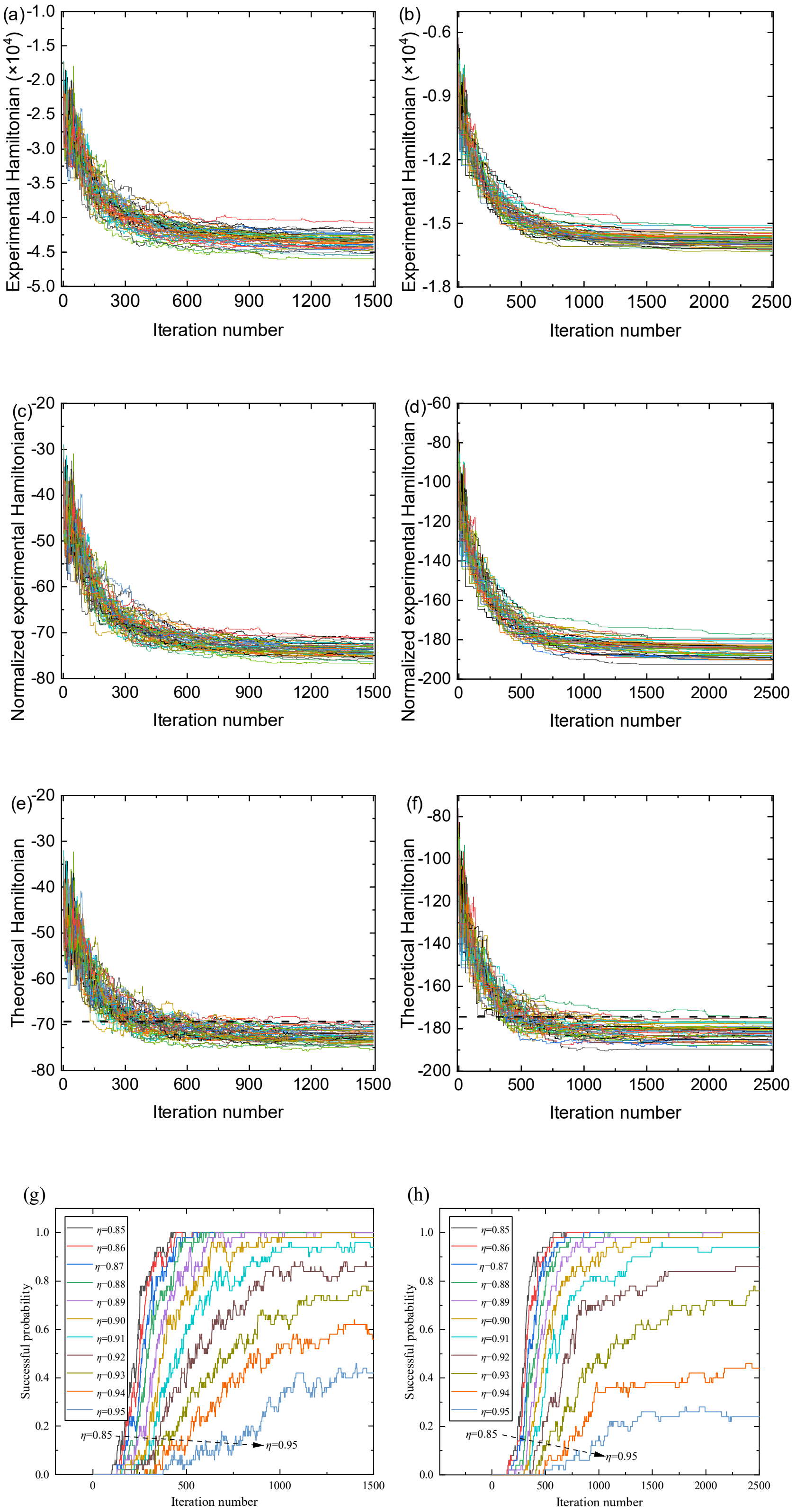

Fig. S6. The normalization of the experimental Hamiltonians and the successful probabilities of the ground state search. (a)-(b) The raw accepted experimental Hamiltonian curves of the 50 runs of model 1-2 respectively. (c)-(d) The normalized accepted experimental Hamiltonian curves of the 50 runs of model 1-2 respectively. (e)-(f) The theoretical accepted Hamiltonian curves of the 50 runs of model 1-2 respectively. The black dashed lines denote the Hamiltonian level when $\eta = 0.9$. (g)-(h) The successful probability curves of model 1-2 respectively.